\documentclass[twocolumn]{autart}    

\usepackage[dvips]{epsfig}    

\begin{document}

\begin{frontmatter}
\runtitle{A Search for Radio Emission from PSR J0537$-$6910}  

\title{A Search for Radio Emission from the Young 16-ms X-ray Pulsar
PSR J0537$-$6910}



\author[froney]{Fronefield Crawford}\ead{fcrawfor@haverford.edu},
\author[maura]{Maura McLaughlin}\ead{mclaughl@jb.man.ac.uk},         
\author[simon]{Simon Johnston}\ead{simonj@physics.usyd.edu.au},
\author[roger]{Roger Romani}\ead{rwr@astro.stanford.edu}, 
\author[froney]{Ethan Sorrelgreen}\ead{esorrelg@haverford.edu} 

\address[froney]{Department of Physics, Haverford College, Haverford,
PA 19041-1392, USA}

\address[maura]{Jodrell Bank Observatory, University of Manchester,
Macclesfield, Cheshire SK11 9DL, UK}

\address[simon]{School of Physics, University of Sydney, NSW 2006,
Australia}

\address[roger]{Department of Physics, 382 Via Pueblo Mall, Stanford,
CA 94305-4060, USA}



\begin{abstract}                         
PSR J0537$-$6910 is a young, energetic, rotation-powered X-ray pulsar
with a spin period of 16 ms located in the Large Magellanic Cloud. We
have searched for previously undetected radio pulsations (both giant
and standard) from this pulsar in a 12-hour observation taken at 1400
MHz with the Parkes 64-m radio telescope.  The very large value of the
magnetic field at the light cylinder radius suggests that this pulsar
might be emitting giant radio pulses like those seen in other pulsars
with similar field strengths.  No radio emission of either kind was
detected from the pulsar, and we have established an upper limit of
$\sim 25$~mJy kpc$^{2}$ for the average 1400-MHz radio luminosity of
PSR J0537$-$6910. The 5$\sigma$ single-pulse detection threshold was
$\sim 750$~mJy for a single 80-$\mu$s sample.  These limits are likely
to be the best obtainable until searches with greatly improved
sensitivity can be made with next-generation radio instruments.
\end{abstract}

\end{frontmatter}

\section{Introduction and Motivation}

\begin{figure}
\begin{center}
\epsfig{file=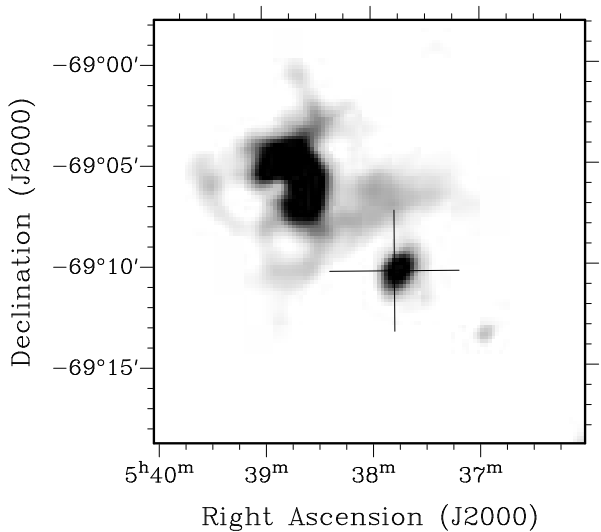,width=8.4cm}
\caption{An 843-MHz radio image of the supernova remnant N157B in the
30 Doradus region of the LMC.  This image was taken from the Sydney
University Molonglo Sky Survey, conducted with the Molonglo
Observatory Synthesis Telescope.  The location of PSR J0537$-$6910, in
the supernova remnant N157B, is indicated by the cross.}
\label{fig1}
\end{center}
\end{figure}

PSR J0537$-$6910 is a fast-spinning, young, and energetic
rotation-powered pulsar that was discovered as a pulsed X-ray source
\cite{mgz+98} in the supernova remnant N157B, located in the Large
Magellanic Cloud (LMC) at a distance of 50 kpc (see Figure
\ref{fig1}). The pulsar is very young, with a characteristic age
$\tau_{c} \equiv P/2\dot{P} \sim 5$~kyr, and its spin-period of 16 ms
makes it the fastest non-recycled pulsar known.  PSR J0537$-$6910 is
one of only a handful of young ($\tau_{c} < 10$ kyr), energetic
pulsars (of which the Crab pulsar is the prototype) that are known. No
radio pulsations were detected in a search \cite{ckm+98} conducted
soon after its discovery in X-rays.

PSR J0537$-$6910 also has the largest inferred dipole magnetic field
strength at the light cylinder radius ($B_{lc}$) of any known
pulsar. The light-cylinder radius is the equatorial distance from the
pulsar at which the co-rotation speed equals the speed of light.  In
Table \ref{tbl-1} we list the top 8 pulsars sorted by decreasing
$B_{lc}$. PSR~J0537$-$6910 is at the top of the list, and the next 7
all show evidence for giant pulse emission. There is no confirmed
evidence of giant pulses from any pulsars with smaller $B_{lc}$
values. Although this may not be an exclusive measure of giant pulse
activity, it certainly appears to be an excellent indicator. We would
therefore expect giant pulses to be emitted by PSR~J0537$-$6910.

Discovery of a radio counterpart to PSR J0537$-$6910 would be
important for several reasons. It could provide a secondary timing
method for the pulsar using ground-based observations.  PSR
J0537$-$6910 is known to glitch frequently \cite{mgm+04}, and radio
timing could be used to study this phenomenon, which is common in
young pulsars \cite{a95}.  A possible determination of the braking
index $n$ from long-term timing might be used to test pulsar spin-down
models \cite{m97}, though this might be impossible given the noisy
timing behavior and frequent glitches observed for the pulsar.  A
measured dispersion measure (DM) for the pulsar would help constrain
the plasma distribution in the LMC \cite{ckm+01}. A radio detection
might also allow a measurement of a phase offset between the X-ray and
radio pulse profiles. This would aid our understanding of the
magnetospheric physics of pulsars \cite{ry95}.  Detection of giant
radio pulses from PSR J0537$-$6910 would confirm the connection
between giant pulses and the light-cylinder magnetic field strength.

\begin{center}
Table 1: Pulsars with the largest values of $B_{lc}$  \\
\end{center}

\begin{center}
\begin{tabular}{lcc} 
\hline
PSR & $B_{lc}$                 & Giant Pulse \\
    & $(\times 10^{5}~\rm{G})$ & Reference \\
\hline
{\it J0537$-$6910}     & {\it 20.6} & -- \\
B1937+21        & 10.2 & \cite{cst+96} \\
B0531+21 (Crab) & 9.8  & \cite{sr68,lcu+95,cbh+04} \\
B1821$-$24      & 7.4  & \cite{rj01} \\
B1957+20        & 3.8  & \cite{jkl+04} \\
B0540$-$69      & 3.7  & \cite{jr03,jrm+04}  \\
J0218+4232      & 3.2  & \cite{jkl+04} \\
B1820$-$30A     & 2.5  & \cite{kbm+05} \\
\hline
\label{tbl-1} 
\end{tabular}
\end{center}

\section{Observations and Analysis}

The Parkes 64-m telescope in Parkes, Australia was used to observe PSR
J0537$-$6910 for a continuous 12-hour observation on 6 September
2003. The center beam of the multibeam receiver \cite{swb+96} was used
at a center observing frequency of 1390 MHz. A 256-MHz bandwidth was
split into 512 contiguous frequency channels, and each channel was
one-bit sampled at 80 $\mu$s.  Data were recorded on magnetic tape at
the observatory and transferred to several sites for processing. The
observing setup was the same as the one used in a recent search for
giant radio pulses from PSR B0540$-$69 \cite{jrm+04}.

\subsection{Standard Pulse Search}

For the standard pulse search, the data were checked for radio
frequency interference (RFI), and a few percent of the data were
subsequently excised. The data were then dedispersed at 75 trial DMs
ranging from 50 to 200 pc cm$^{-3}$, corresponding to the expected DM
range for LMC pulsars \cite{ckm+01}. A DM trial spacing of 2 pc
cm$^{-3}$ was chosen to ensure that the pulse smear from dedispersion
error did not exceed 5\% of the pulse period.\footnote{This is
comparable to the (uncorrectable) intra-channel smearing in the system
for a DM of about 100 pc cm$^{-3}$.}

For each DM trial, the data were analyzed using both a standard Fast
Fourier Transform (FFT) search of the dedispersed time series
\cite{rem02} and a routine which folded each dedispersed time series
at a range of periods around the nominal period derived from the X-ray
ephemeris \cite{mgz+98}.  Since PSR J0537$-$6910 was observed to
glitch six times in a 2.6-year span with an average glitch magnitude
of $\Delta P / P \sim 0.4 \times 10^{-6}$ \cite{mgm+04}, the true
period is probably somewhat offset from the ephemeris period. Periods
$\pm 1000$ ns from the ephemeris period were tried, with a trial fold
step size of 0.5 ns. This again ensured that the pulse smear from
folding at the wrong period would not exceed 5\% of the pulse period.

\subsection{Giant Pulse Search}

For the giant pulse search, the data were dedispersed at 1000 trial
DMs ranging from 0 to 300 pc cm$^{-3}$. This narrow DM spacing ensured
that the smearing due to an incorrect DM was less than $\sim 0.1$ ms.
Two different software codes were used on the data. The algorithms for
both are similar and are described in detail in \cite{mc03} and
\cite{rj01}. In brief, each dedispersed time series was searched for
single pulses above a 5$\sigma$ significance threshold. The time
series were smoothed multiple times by aggregating adjacent samples in
groups of 2, 4, 8, etc., and the search was repeated on each smoothed
time series. This technique increased sensitivity to broadened
pulses. A procedure was incorporated to remove pulses strongest at low
DMs and therefore most likely due to RFI.

\section{Results and Conclusions} 

No significant radio signal was detected in either the FFT search or
the folding search. No FFT candidates above a signal-to-noise
threshold of 5 and no folded profiles with significantly large
$\chi^{2}$-values were found. No individual pulses with high DMs and
signal-to-noise ratios greater than 7 were detected and no excess of
weaker pulses was detected at a specific DM. The 5$\sigma$
single-pulse detection threshold was $\sim 750$~mJy for a single
80-$\mu$s sample.  For comparison, previous observations of giant
pulses from the Crab pulsar \cite{cbh+04} indicate that if the Crab
were located in the LMC, we would expect to detect one giant pulse
every $\sim 20$ minutes from this source at 1400 MHz; the strongest
pulse during a 12-hour observation of this source would have a
signal-to-noise ratio of $\sim 40$. For PSR B0540$-$69, one giant
pulse is detected every $\sim 30$ minutes with a similar observing
setup. Giant pulses from PSR J0537$-$6910 must therefore be a least a
factor of two weaker than those from PSR B0540$-$69.

\begin{center}
Table 2: Estimated radio luminosities of rotation-powered 
pulsars with $\tau_{c} < 10$ kyr \\
\end{center}

\begin{center}
\begin{tabular}{lccc} 
\hline
PSR & $\tau_{c}$    & $P$   & $L_{1400}$  \\
    & (kyr)         & (sec) & (mJy kpc$^{2}$)       \\
\hline
J1846$-$0258      & 0.72 & 0.324 &              $< 50$     \\
B0531+21 (Crab)   & 1.24 & 0.033 &                 56      \\
B1509$-$58        & 1.55 & 0.150 &                 27      \\ 
J1119$-$6127      & 1.61 & 0.407 &                 20      \\
B0540$-$69        & 1.67 & 0.050 &                 60      \\
J1124$-$5916      & 2.87 & 0.135 &                  2.3    \\
J1930+1852        & 2.89 & 0.137 &                  1.5    \\
{\it J0537$-$6910}      & {\it 4.98} & {\it 0.016} &        {\it $<$ 25}  \\
J0205+6449        & 5.37 & 0.065 &                  0.46   \\  
J1357$-$6429      & 7.30 & 0.166 &                  2.7    \\
J1614$-$5048      & 7.42 & 0.232 &                130      \\
J1617$-$5055      & 8.13 & 0.069 &                 20      \\
J1734$-$3333      & 8.13 & 1.169 &                 27      \\
\hline
\label{tbl-2} 
\end{tabular}
\end{center}

An upper limit for the 1400-MHz flux density of PSR J0537$-$6910 was
determined using the radiometer equation with an additional factor to
account for pulsed duty cycle \cite{dtw+85}. For an assumed duty cycle
of 5\%, the sensitivity limit from the observation was determined to
be $S_{1400}^{\rm min} \sim 10~\mu$Jy. This translates into an average
1400-MHz radio luminosity upper limit of $L_{1400}^{\rm min} =
S_{1400}^{\rm min} d^{2} \sim 25$ mJy kpc$^{2}$ (where $d \sim 50$ kpc
is the pulsar's distance). This limit is comparable to or less than
the luminosities of several pulsars with $\tau_{c} < 10$
kyr.\footnote{In some cases, the true age of the pulsar is believed to
be significantly different than its characteristic age, $\tau_{c}$.}
Table 2 
shows the list of known rotation-powered pulsars
with $\tau_{c} < 10$ kyr, rank-ordered by increasing $\tau{_c}$. As
can be seen from the table, the luminosity range for the young pulsar
population spans more than two orders of magnitude, so no strict
conclusions can be made about whether PSR J0537$-$6910 is actually a
radio emitter. It may be the case that more sensitive observations in
the future will be able to detect radio emission from PSR
J0537$-$6910, but the limits presented here are likely to be the best
obtainable for the foreseeable future. It may also be that PSR
J0537$-$6910 is indeed a strong radio emitter, but that its radio beam
is misaligned with our line of sight.  With the availability of
next-generation radio instruments, such as the Square Kilometer Array,
observations of PSR J0537$-$6910 with greatly improved sensitivity
will be possible.

\begin{ack}                          
This work was funded in part by the Keck Northeast Astronomy
Consortium, the American Astronomical Society International Travel
Grants Program, the Haverford College Green Research Fund, and REU
supplement AST-0330701 to NSF grant AST-0071192.
\end{ack}

\bibliographystyle{plain}        
\bibliography{cmj+05}            



\end{document}